\documentclass[letter]{aa}

\pdfoutput=1

\usepackage{graphicx}
\usepackage{amssymb}
\usepackage{epstopdf}
\usepackage{isolatin1} 		
\usepackage{url}
\usepackage{hyperref}
\usepackage{natbib}
\bibpunct{(}{)}{;}{a}{}{,} 		
\usepackage{txfonts} 		

\hypersetup{
    colorlinks=false,
    urlcolor=blue,
    bookmarks,
    bookmarksnumbered,
    pdftitle={Hard X-ray flares in IGR J08408-4503 unveil clumpy stellar winds},
    pdfauthor={Leyder J.-C., Walter R., Lazos M., Masetti N. and Produit N.},
    pdfsubject={Hard X-ray flares in IGR J08408-4503 unveil clumpy stellar winds},
    pdfkeywords={gamma rays: observations  - stars: supergiants  - X-rays: individuals:  IGR~J08408-4503 - stars: early-type - instabilities},
    pdfcenterwindow=true
}

\begin{document}

\title{Hard X-ray flares in \object{IGR~J08408--4503} unveil clumpy stellar winds}

\author{
J.-C. Leyder \inst{1,2} \fnmsep\thanks{FNRS Research Fellow} \and
R. Walter \inst{2,3} \and
M. Lazos \inst{2,4} \and
N. Masetti \inst{5} \and
N. Produit \inst{2,3}
}

\offprints{J.-C. Leyder}

\institute{
Institut d'Astrophysique et de Géophysique, Université de Liège, Allée du 6-Août 17, Bâtiment B5c, B--4000 Liège, Belgium \\ \email{leyder@astro.ulg.ac.be} \and
\textit{INTEGRAL} Science Data Centre, Université de Genève, Chemin d'Écogia 16, CH--1290 Versoix, Switzerland \and
Observatoire de Genève, Université de Genève, Chemin des Maillettes 51, CH--1290 Sauverny, Switzerland \and
Centro de Astrofísica da Universidade do Porto, Rua das Estrelas, PT--4150-762 Porto, Portugal \and
INAF --- Istituto di Astrofisica Spaziale e Fisica Cosmica di Bologna, Via Gobetti 101, I--40129 Bologna, Italy (formerly IASF/CNR, Bologna)
}

\date{Received 29 August 2006 / Accepted 14 February 2007}

\abstract
{A 1000-s flare from a new hard X-ray transient, \object{IGR~J08408--4503}, was observed by \textit{INTEGRAL} on May 15, 2006 during the real-time routine monitoring of IBIS/ISGRI  images performed at the \textit{INTEGRAL} Science Data Centre. The flare, detected during a single one-hour long pointing, peaked at 250~mCrab in the 20--40~keV energy range.
}
{Multi-wavelength observations, combining high-energy and optical data, were used to unveil the nature of \object{IGR~J08408--4503}.
}
{A search in all \textit{INTEGRAL} public data for other bursts from \object{IGR~J08408--4503} was performed, and the detailed analysis of another major flare is presented. The results of two \textit{Swift} Target of Opportunity observations are also described. Finally, a study of the likely optical counterpart, \object{HD~74\,194}, is provided.
}
{\object{IGR~J08408--4503} is very likely a supergiant fast X-ray transient (SFXT) system. The system parameters indicate that the X-ray flares are probably related to the accretion of wind clumps on a compact object orbiting about $10^{13}$~cm from the supergiant \object{HD~74\,194}. The clump mass loss rate is of the order of $10^{-6}$~$M_{\sun}$\,yr$^{-1}$.
}
{Hard X-ray flares from SFXTs allow to probe the stellar winds of massive stars, and could possibly be associated with wind perturbations due to line-driven instabilities.
}



\maketitle

\section{Introduction}
\label{sec:Introduction}
A new hard X-ray transient, \object{IGR~J08408--4503}, was discovered by \textit{INTEGRAL} on May 15, 2006 during the real-time routine monitoring of IBIS/ISGRI 20--40~keV images performed at the \textit{INTEGRAL} Science Data Centre (ISDC). It was detected during a one-hour long pointing, and further analysis of the lightcurve with a 100-s binning showed that the source flared for $\sim$1000~s starting at 2006-05-15T18:25:33, with a peak flux of 250 mCrab in the 20--40~keV energy range \citep{Gotz+06}. The source was also detected by the JEM-X instrument, providing a position of~: RA(J2000)~= $08^\mathrm{h}40^\mathrm{m}48\,\fs7$, Dec(J2000)~= $-45\degr03\arcmin41\arcsec$ ($\pm46\arcsec$ at 90\% confidence) \citep{Brandt+06}.

Analysis of the public \textit{INTEGRAL} archive shows that there was at least another outburst from this source on July 1, 2003 \citep{Mereghetti+06}. 
A detailed analysis of this first flare is presented in Sect.~\ref{subsec:Analysis-INTEGRAL}.

A first \textit{Swift} target of opportunity observation was carried out on \object{IGR~J08408--4503} on May 22, 2006 \citep{Kennea+06}. The XRT observation (in photon counting mode) revealed the presence of a faint source in the JEM-X error box at the position~: RA(J2000)~= $08^\mathrm{h}40^\mathrm{m}48^\mathrm{m}$, Dec(J2000)~= $-45\degr03\arcmin30\arcsec$ (with an error radius of $5\,\farcs4$ at 90\% confidence). Sect.~\ref{subsec:Analysis-Swift} gives more details about this observation and a following one.

Preliminary studies of ESO archival optical spectra of the star \object{HD~74\,194}, which is lying at the \textit{Swift}/XRT position, were performed by \citet{Masetti+06ATel}, who pointed out that this star is likely associated with \object{IGR~J08408--4503}, on the basis of the detection of H$\alpha$ in emission. A deeper analysis of these results is presented in Sect.~\ref{sec:Analysis-ESO}.

Finally, Sect.~\ref{sec:Discussion} provides a discussion on the nature of this new source, and derives likely system parameters based on the observations.

\section{The high-energy source}
\label{sec:High-Energy}
\subsection{\textit{INTEGRAL} observations}
\label{subsec:Analysis-INTEGRAL}
All available public \textit{INTEGRAL} \citep{WInkler+03} pointings less than 10\degr\ away from the source position were analysed to look for other flares from the source. There are two major periods during which the source was frequently observed, summarized in Table~\ref{tab:INTEGRAL-data}. The data were analyzed using OSA\footnote{The offline scientific analysis (OSA) software is available from the ISDC website~: \mbox{\url{http://isdc.unige.ch}}}, version 5.1. A lightcurve at the source position was extracted, with a binning of 100~s.

\begin{table}[htdp]
\caption{Summary of the main \textit{INTEGRAL} public data available for \object{IGR~J08408--4503}.}
\label{tab:INTEGRAL-data}
\centering
\begin{tabular}{ccc} \hline \hline
Rev. & Period [MJD] & Date \\ \hline 
81--88 & 52\,803--52\,827 & 12 Jun--6 Jul 2003 \\
137--141 & 52\,970--52\,984 & 27 Nov--11 Dec 2003 \\ \hline
\end{tabular}
\end{table}

\begin{figure}[htbp]
\centering
\includegraphics[width=0.75\columnwidth]{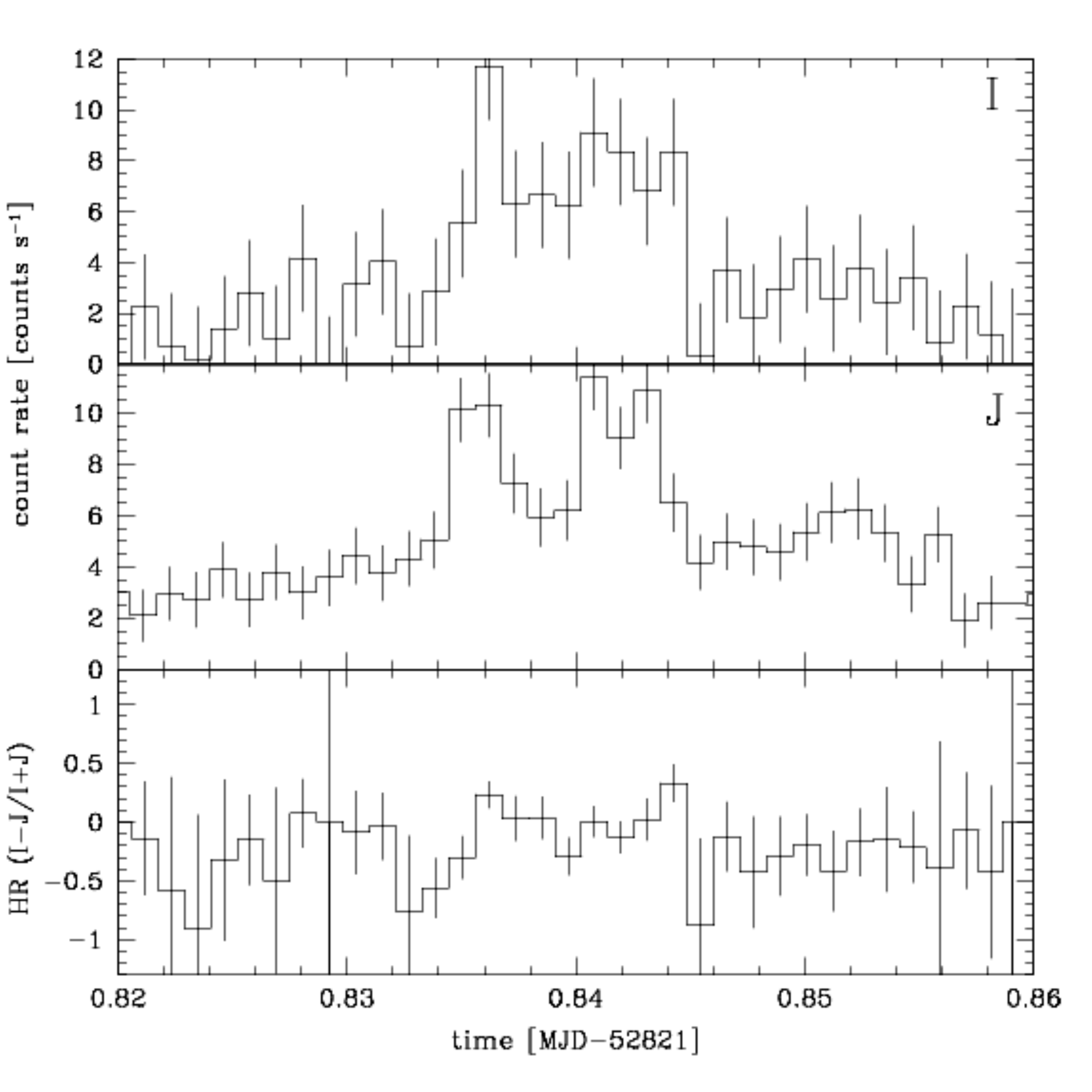}
\caption{ISGRI (I, 20--60~keV) and JEM-X (J, 3--35~keV) lightcurves around the flare of \object{IGR~J08408--4503} (with a 100~s binning), and hardness ratio (I-J)/(I+J).}
\label{fig:flare-lightcurve}
\end{figure}

The detailed study of this lightcurve showed that there is only one significant flare, around 2003-07-01T20:03:02. A zoom on the ISGRI and JEM-X lightcurves around this flare is shown in Fig.~\ref{fig:flare-lightcurve}, together with the hardness ratio. The source was significantly detected with ISGRI for 12.6~ks with a maximum flux reaching 11.7~counts\,s$^{-1}$ in the 20--60~keV energy band in the 100-s bin centered around 2003-07-01T20:03:02.552 (in pointing \texttt{00870035}). Meanwhile, the JEM-X flux peaked at 11.4~counts\,s$^{-1}$ in the 3--35~keV energy band.

The peak 2--10~keV X-ray flux could be estimated as $1.1 \times 10^{-9}$~erg\,s$^{-1}$\,cm$^{-2}$, assuming the spectral model described below (\texttt{wabs}*\texttt{comptt}). The average flux during the 1000-s flare is $7 \times 10^{-10}$~erg\,s$^{-1}$\,cm$^{-2}$.

Mosaicking 6 pointings surrounding the main flare (3 before and 3 after, \textit{i.e.} excluding the pointing with the flare itself) shows that \object{IGR~J08408--4503} is also clearly detected around the main flare, from 2003-07-01T16:40:00 until 23:35:13.

A mosaic of all pointing images containing the source (except 8 pointings surrounding the flare) was produced and gave a 5-$\sigma$ upper limit for the flux in quiescence of 0.056~counts\,s$^{-1}$ in the 23--40~keV energy range, for an effective exposure time of 2.45~Ms. The 2--10~keV upper limit can be estimated as $6 \times 10^{-12}$~erg\,s$^{-1}$\,cm$^{-2}$.

\begin{figure}[htbp]
\centering
\includegraphics[width=0.8\columnwidth]{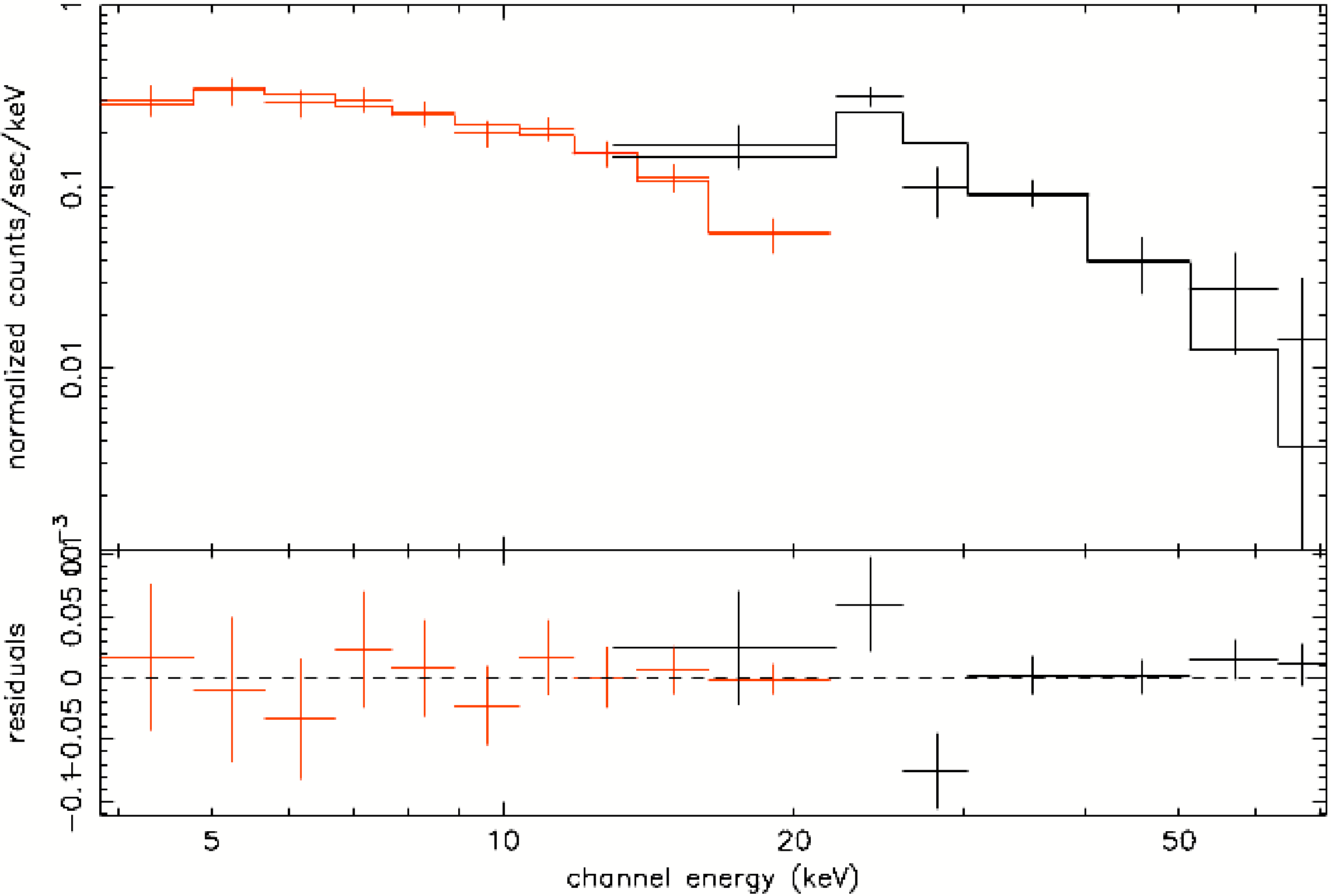}
\caption{JEM-X and ISGRI spectra of \object{IGR~J08408--4503} (for pointing 35 only), fitted with a \texttt{constant*wabs*comptt} model.}
\label{fig:spectra-1ScW}
\end{figure}

The JEM-X and ISGRI average spectra of \object{IGR~J08408--4503} were extracted from the pointing containing the flare, and are shown in Fig.~\ref{fig:spectra-1ScW}. The spectrum was fitted\footnote{For all fits, a \texttt{constant} is added to allow for normalisation between JEM-X and ISGRI, and a \texttt{wabs} is added to take into account the hydrogen column, $N_{H}$.} with \texttt{xspec} using a simple absorbed \texttt{powerlaw}, but no good fit could be obtained with $\chi_{\nu}^2 \simeq 1.7$ (for 13 degrees of freedom). Then, two models exhibiting a spectral break were used~: a broken powerlaw (\texttt{bknpower}) and a comptonisation model (\texttt{comptt}). Both models give much better results, with the $\chi_{\nu}^2$ for the \texttt{comptt} model decreasing to 1.0 (for 12 degrees of freedom). Applying an \texttt{F-test} shows that there is a 99.3\% probability that the \texttt{comptt} model is better than the simple \texttt{powerlaw}, implying that a spectral break is needed. The \texttt{bknpower} model suggests a break at an energy of $14.8\pm3.4$~keV. The \texttt{comptt} model gives a plasma temperature $kT$ of $7\pm1$~keV, and a plasma optical depth $\tau_{\mathrm{plasma}}$ of $4\pm1$. The intercalibration constant between ISGRI and JEM-X is $1.2\pm0.3$ in the case of the \texttt{powerlaw} model, and $1.3\pm0.3$ for the \texttt{comptt} model.

Only an upper limit on the value of the hydrogen column $N_{H}$ can be obtained~: the 2-$\sigma$ upper limit on $N_{H}$ is $3 \times 10^{23}$~cm$^{-2}$.

\subsection{\textit{Swift}/XRT observations}
\label{subsec:Analysis-Swift}
Two \textit{Swift} \citep{Gehrels+04} Target of Opportunity observations of \object{IGR~J08408--4503} were performed~: they consisted of a nominal 3.6 and 4.0~ks exposure in photon counting mode starting at 2006-05-22T10:41:01 and at 2006-07-29T00:15:00. The data were analysed with version 2.3 of the \textit{Swift} software\footnote{The \textit{Swift} software is available from the GSFC website~: \mbox{\url{http://swift.gsfc.nasa.gov/}}}.

The best position for \object{IGR~J08408--4503} (derived from the second observation) is~: RA(J2000)~= $08^\mathrm{h}40^\mathrm{m}47^\mathrm{s}$, Dec(J2000)~= $-45\degr03\arcmin32\arcsec$ (with an uncertainty of $4\,\farcs1$ at 90\% confidence level), consistent with the results of \citet{Kennea+06}.

The source is very faint, with only 22~counts in total (resp. 20 for the second observation), and $18.3\pm4.7$~counts after background subtraction (resp. $18.4\pm4.5$). Assuming the spectral model derived from \textit{INTEGRAL} (\texttt{wabs}*\texttt{comptt}), the flux measured by \textit{Swift} can be estimated as $3.4 \times 10^{-13}$~erg\,s$^{-1}$\,cm$^{-2}$ in the 2--10~keV energy band (resp. $3.1 \times 10^{-13}$).
However, this model is not compatible with the fact that all (but one) events detected by \textit{Swift} have an energy smaller than 1.5~keV. Assuming a soft thermal model, one find that these 22 counts are compatible with a 0.5--2~keV flux of $9.4 \times 10^{-14}$~erg\,s$^{-1}$\,cm$^{-2}$.

\section{Optical counterpart}
\label{sec:Analysis-ESO}
\subsection{Imaging}
\label{subsec:Imaging}
\begin{figure}[htbp]
\centering
\includegraphics[width=0.8\columnwidth]{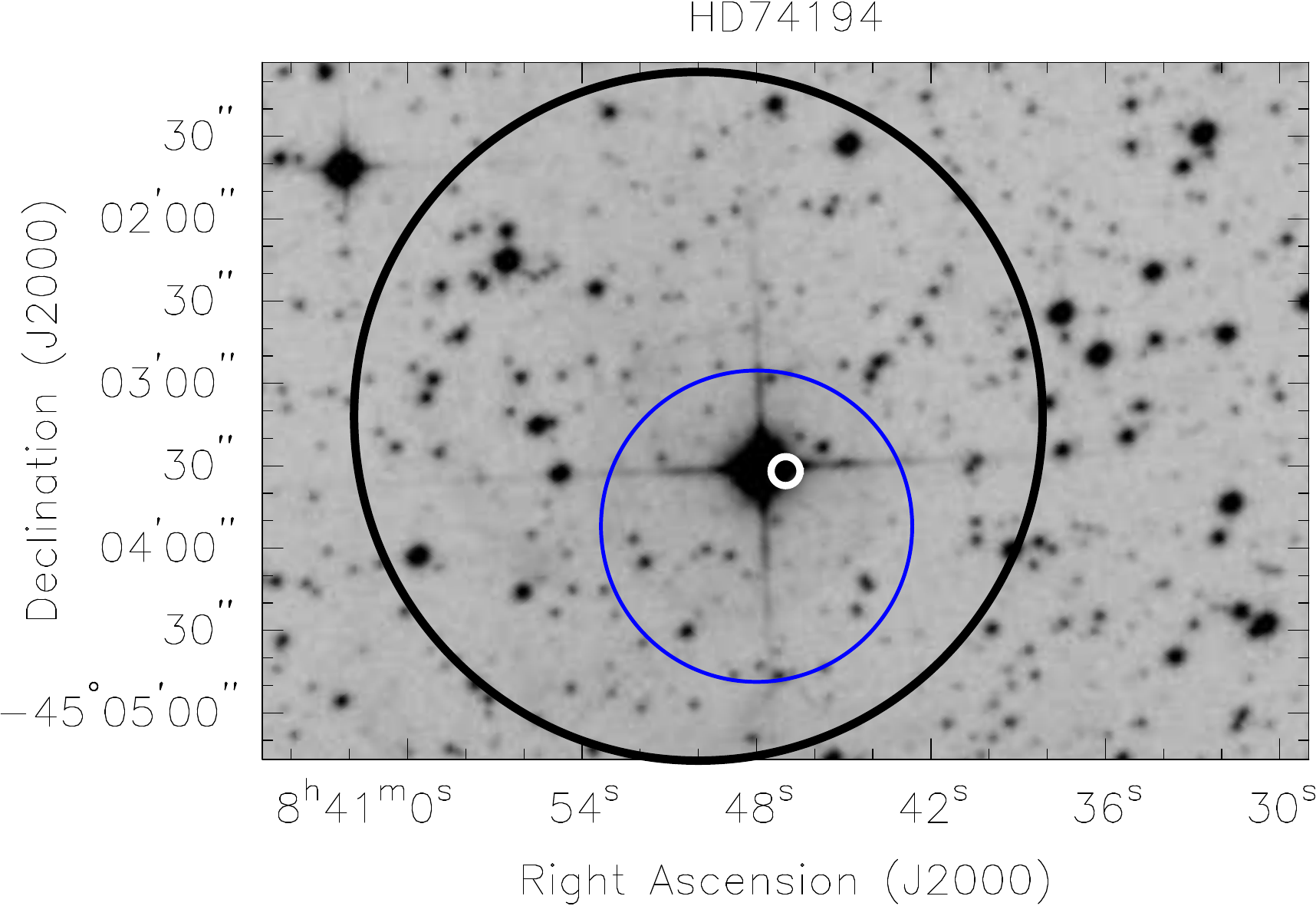}
\caption{DSS-II-Red optical image of the field of \object{IGR~J08408--4503}. The outer black and middle blue circles indicate respectively the \textit{INTEGRAL} ISGRI and JEM-X error circles, and the innermost white circle indicates the \textit{Swift} XRT position. The star \object{HD~74\,194} matches perfectly. The field size is $5.8\arcmin \times 3.7\arcmin$; North is up and East to the left.}
\label{fig:image-dss}
\end{figure}

A DSS-II red image of the field is shown in Fig.~\ref{fig:image-dss}. In the 2MASS catalogue, there is only one star in the \textit{Swift} error circle~: \object{HD~74\,194}, which has a position of~: RA(J2000)~= $08^\mathrm{h}40^\mathrm{m}47\,\fs79$, Dec(J2000)~= $-45\degr03\arcmin30\,\farcs2$. The next closest source is nearly 10'' away from \object{IGR~J08408--4503}. However, \object{HD~74\,194} is so bright that we cannot exclude the presence of other faint sources in the error circle. Therefore, we cannot conclusively associate \object{IGR~J08408--4503} with \object{HD~74\,194} by positional arguments alone.

\subsection{Spectroscopy}
\label{subsec:Spectroscopy}
Spectroscopy of the star \object{HD~74\,194} was secured under programme \texttt{70.C-0396(A)} on January 9, 2003 at the ESO 3.6m telescope equipped with EFOSC, which uses a (2$\times$2 binned) 2060$\times$2060 pixel Loral/Lesser CCD. Two 5-s spectra were acquired starting at 08:38:39 UT; grism \#12 and a slit width of $1''$ were used, providing a 6\,000--10\,000 \AA~nominal spectral coverage. The use of this setup guaranteed a final dispersion of 4.2~\AA/pix.

After cosmic-ray rejection, the spectra were reduced, background subtracted and optimally extracted \citep{Horne+86} using IRAF\footnote{The image reduction and analysis facility (IRAF) software is available from the NOAO website~: \mbox{\url{http://iraf.noao.edu/}}}. Wavelength calibration was performed using He-Ar lamps. The spectra were then simply normalized to their continuum because no spectroscopic standard was available for the star \object{HD~74\,194} in the ESO archive, and were stacked together to increase the S/N ratio. The wavelength calibration uncertainty was $\sim$0.5~\AA~for all cases; this was checked using the positions of background night sky lines.

\begin{figure}[htbp]
\centering
\includegraphics[height=0.5\columnwidth, angle=270]{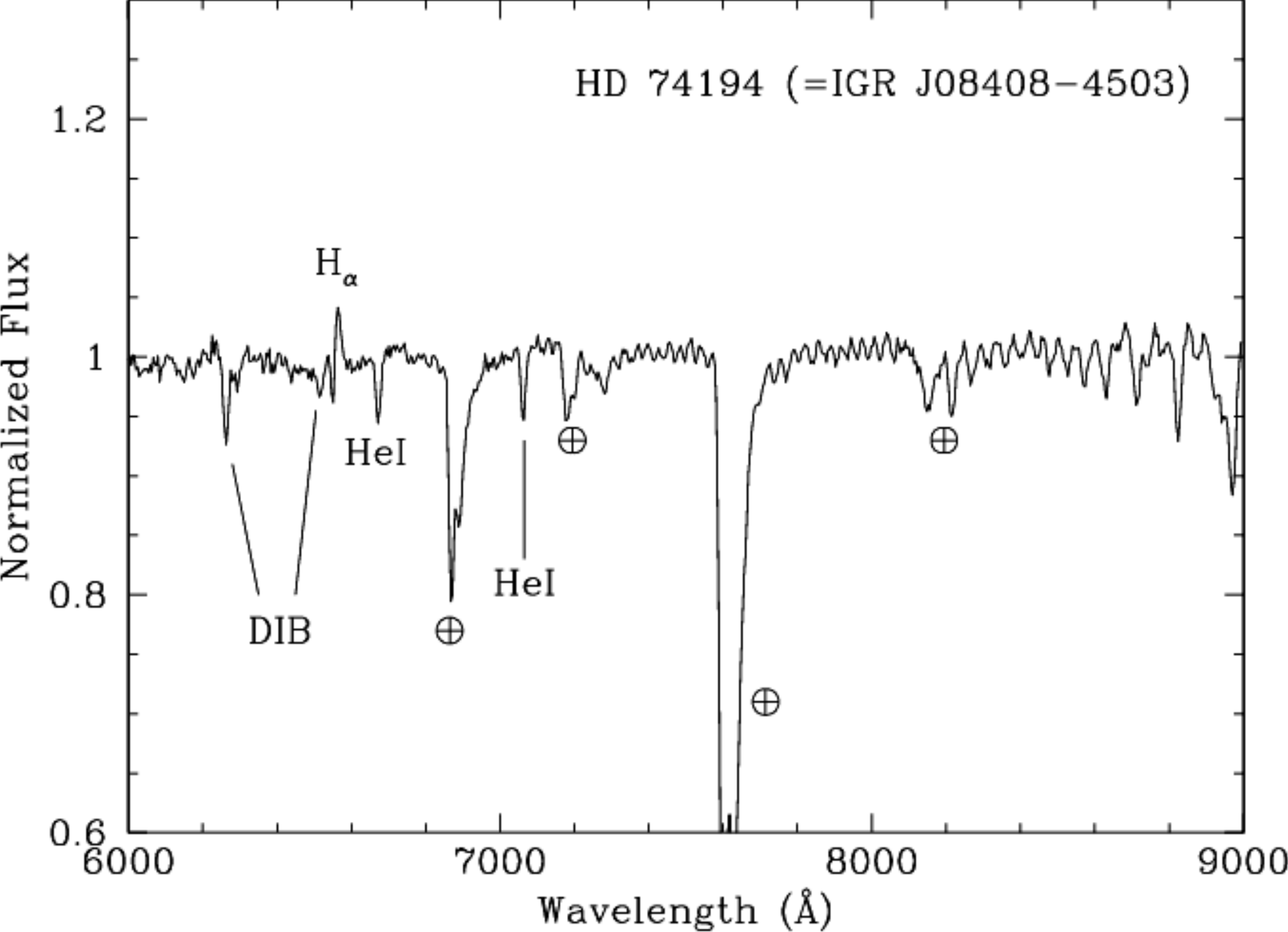}
\includegraphics*[height=0.49\columnwidth, angle=270]{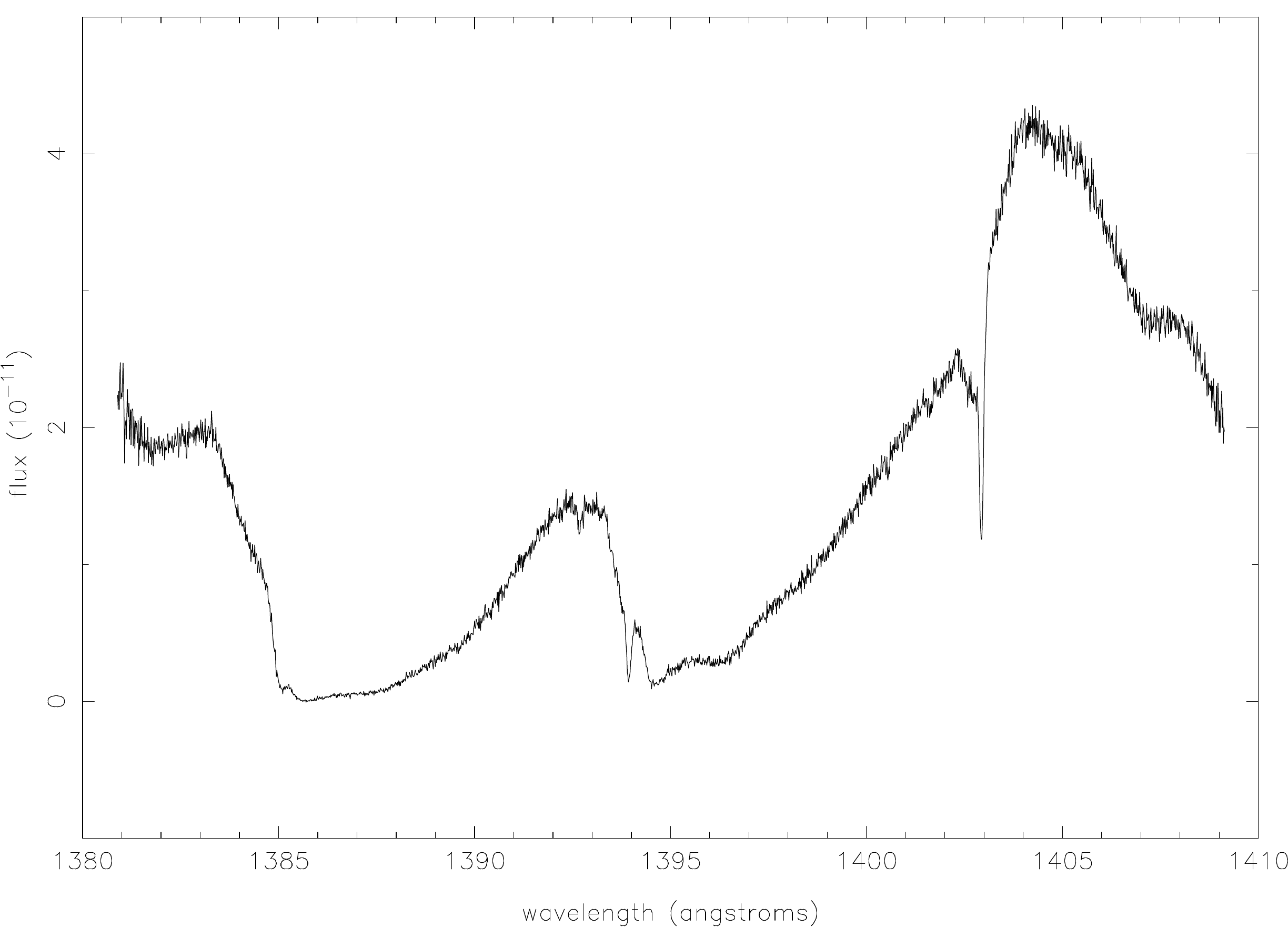}
\caption{\textit{Left~:} Combined 6000--9000 \AA~spectrum of the star \object{HD~74\,194}, acquired with the 3.6m ESO telescope. The main spectral features are labeled. The symbol $\oplus$ indicates atmospheric telluric absorption bands. \textit{Right~:} \textit{HST}/GHRS 1380--1410~\AA\ spectrum of \object{HD~74\,194}, showing the \ion{S}{iv} and \ion{O}{iv} lines (dataset \texttt{Z37J020DT}).}
\label{fig:spectra-optical}
\end{figure}

The analysis of the optical spectrum of \object{HD~74\,194} (Fig.~\ref{fig:spectra-optical}) shows the H$\alpha$ line in emission (with EW = 1.0$\pm$0.1 \AA) with a P-Cyg profile; besides, \ion{He}{i} absorption lines at 6\,675 and 7\,065~\AA\ and the diffuse interstellar bands (DIBs) at 6\,280~\AA\ and 6\,515~\AA\ are also detected. 

The P-Cyg profile indicates an escape velocity of $600\pm50$~km\,s$^{-1}$.
Assuming that $v_{\infty}/v_{\mathrm{escape}} = 2.78\pm0.36$ \citep{Groenewegen+89}, this is consistent with the value of $v_{\infty} = 2000$~km\,s$^{-1}$ derived from \textit{IUE} spectra by \citet{Lamers+95} and \citet{Snow+94}. We also analysed the archival \textit{HST}/GHRS spectrum of \object{HD~74\,194} around the \ion{O}{iv} and \ion{Si}{iv} lines, and found maximum velocities of $3100\pm100$~km\,s$^{-1}$ for both lines (see Fig.~\ref{fig:spectra-optical}).

The positional consistency of this star with the quiescent emission of \object{IGR~J08408--4503} suggests that \object{HD~74\,194} is an O-type supergiant fast X-ray transient.

The fact that \object{HD~74\,194} is of spectral type O8.5\,Ib \citep{Walborn+73} implies M$_V$ = $-$6.2 and $(B-V)_0$ = $-$0.285 \citep{Lang+92}. Using the observed $V$ = 7.55 and $B-V$ = +0.22 for this star \citep{Drilling+91}, we obtain a color excess of $E(B-V)$ = +0.505 and a distance $d \sim 2.7$~kpc. 
This distance estimate is in agreement with that obtained from the \textit{Hipparcos} satellite \citep{Perryman+97}, which measured an annual parallax $\pi = 0.36\pm0.64$~milliarcsec for \object{HD~74\,194}; this value implies a distance $d \sim 2.8$~kpc (with a large error) to this source. 
We also estimated the color excess from \textit{IUE} archival spectra (datasets \texttt{LWP2966[1-2]} and \texttt{SWP5302[3-5]}), by calculating the reddening necessary to compensate for the 2200~\AA\ dust extinction feature, and found $E(B-V)$ = +0.63, similar to the value derived above when taking into account the uncertainty on the extinction law.
The color excess is less than half of the value of +1.66 obtained from infrared maps \citep{Schlegel+98}, thereby supporting a relatively close distance to Earth. This distance places \object{HD~74\,194} in the Carina Galactic spiral arm tangent \citep[see, e.g.,][]{Leitch+98},
and in this direction, the interstellar absorption column density can be estimated as $N_{H}^{\mathrm{ISM}} \simeq 3 \times 10^{21}$~cm$^{-2}$.

\section{Discussion}
\label{sec:Discussion}
Fast hard X-ray outbursts are mostly known in three broad classes of objects discussed below~: Be transients, thermonuclear flashes, and supergiant fast X-ray transient (SFXT) systems. 

\subsection{Be transient systems}
\label{subsec:A0538}
The outbursts exhibited by most Be transient systems are of 2 different origins : they are either related to the orbital period of the neutron star reaching the disk of the Be star; or they are large, seemingly random, and last a few days.

Few Be HMXBs are known to show fast and occasional outbursts. An example is \object{A0538--66}, which is probably a super-Eddington X-ray source \citep{Skinner+82}.

The flares of \object{IGR~J08408--4503} are probably not of this type, because the timescales are completely different, its flaring luminosity is less than 1\% of the Eddington luminosity, and its optical counter-part is not a Be-type star.

\subsection{Thermonuclear flashes}
\label{subsec:LMXBs}
Alternatively, thermonuclear flashes -- similar to those seen in LMXBs -- could be envisioned. But in HMXBs, the presence of a magnetic field funnels the accretion stream onto the magnetic polar caps of the neutron star (NS), thus preventing the formation of a boundary layer in which H and He could burn. Moreover, these type I flashes are shorter ($\simeq$ 30~s), have a much higher peak luminosity, and are much softer than observed in \object{IGR~J08408--4503} \citep{Remillard+06}. Therefore, the origin of the flares in this case is most likely not linked to thermonuclear flashes.

\subsection{Supergiant Fast X-ray Transients}
\label{subsec:SFXTs}
\object{IGR~J08408--4503} shares many characteristics with other SFXT systems discovered by \textit{INTEGRAL} \citep[e.g.][]{Sguera+06}~:
\begin{itemize}
\item two X-ray flares have been observed from \object{IGR~J08408--4503} in 12 weeks of cumulated \textit{INTEGRAL} observations;
\item the bright flares are fast, with a typical duration\footnote{The total duration of the first flare of \object{IGR~J08408--4503} observed by \textit{INTEGRAL} was at least 6~hours.} of 1000~s;
\item the X-ray luminosity during the 1000~s flare is $6 \times 10^{35}$~erg\,s$^{-1}$, while the quiescence luminosity is lower than $5 \times 10^{33}$~erg\,s$^{-1}$ and has been as low as $3.0 \times 10^{32}$~erg\,s$^{-1}$ during the \textit{Swift}/XRT observation;
\item during flares, the hard X-ray spectrum features a cutoff at $14.8\pm3.4$~keV that is typical of accreting pulsars;
\item the optical counterpart is a supergiant OB star.
\end{itemize}
It is impossible to formally exclude that the X-ray source may be unrelated to the O star. However, for the rest of the discussion, the many similarities between \object{IGR~J08408--4503} and other SFXTs will be taken as an evidence for its association with \object{HD~74\,194}.

The typical X-ray luminosity of O-type stars ranges from $10^{31}$ to $10^{34}$~erg\,s$^{-1}$, so it is likely that the X-ray luminosity observed in quiescence ($3 \times10^{32}$~erg\,s$^{-1}$) is emitted either by the star itself (through shocks in its wind), or by the interaction of its wind with the interstellar medium.
The very soft spectrum observed in quiescence also does not show any evidence for accretion.

\object{HD~74\,194} features micro-variability with a period of 7.8~days \citep{Koen+02}. Such a micro-variability has been associated with stellar spots, or with density waves in the atmosphere \citep{vanGenderen+89}. Nevertheless, the variability amplitude of 5\% does not seem to be related to the observed X-ray variability and is too large to be the signature of an orbital period. The high luminosity X-ray flares, with their hard spectra, are to be explained by another mechanism.

The stellar wind parameters of \object{HD~74\,194}, in particular its mass loss rate $\dot{M}=1.4 \times 10^{-6}$~$M_{\sun}$\,yr$^{-1}$ \citep{Krticka+01} and its terminal velocity $v_{\infty}=2000\pm300$~km\,s$^{-1}$ \citep{Lamers+95}, can be used as input for different models.

If a NS orbits within the wind of an O-type star, an X-ray luminosity of $10^{36} (r_{\mathrm{orb}}/10^{12})^{-2}$~erg\,s$^{-1}$ (where $r_{\mathrm{orb}}$ is the orbital radius in cm) is expected, which is much higher than observed on average in the present case, unless $r_{\mathrm{orb}}$ is larger than $10^{13}$~cm.

The flip-flop instability of wind accretion flow \citep{Benensohn+97} cannot account for the observed variability~: it predicts variations of the accretion rate by a factor up to 10 only, \textit{i.e.} much smaller than observed, and on a time scale $R_{\mathrm{a}}/v_{\infty} \simeq 10^2 - 10^3$~s (where $R_{\mathrm{a}}$ is the effective accretion radius), much too short to explain the observed flare frequency.

The propeller mechanism \citep{Illarionov+75} could also be envisioned to explain the outbursts, supposing that during quiescence, the NS is rotating fast enough to centrifugally inhibit the accretion of matter from the O-type companion. Outbursts would then be observed when a change occurs, switching off the propeller mechanism for a short period. Considering the observed burst duration, the changes cannot be driven by rotation. It is also very unlikely that the ignition and stop of accretion could be driven by a sudden change of the magnetic field and back within a few~ks.

The most likely source for the observed flaring behavior is therefore a sudden change of  the density of the gas trapped within the accretion radius~: this is expected to happen if the stellar wind is clumpy.

Assuming that the flares are the signature of the accretion of wind clumps on a compact object orbiting at $10^{13}$~cm, and that the flare duration is related to the clump size, the clump  column density can be evaluated as $5 \times10^{23}$~cm$^{-2}$. Assuming a spherical shape for the clumps, their typical mass can be estimated as $10^{18}$~g. The flare frequency can then be used to estimate the stellar mass loss rate in the form of clumps as $10^{-6}$~$M_{\sun}$\,yr$^{-1}$ (for a spherical geometry).

The density of the clumps ($\sim10^{13}$~g\,cm$^{-3}$) is only slightly above the peak densities predicted by hydrodynamic simulations of growing wind perturbations induced by line-driven instabilities \citep[see Fig.~6 in][]{Feldmeier+97}. The predicted density ratio of $10^{4}$--$10^{6}$ between the dense and hot phases of the wind at a given radius may explain the large variations between the flaring and quiescent luminosities.

The wind clumps are massive enough to generate the outbursts while being rare enough to explain their low frequency.
Using the observed luminosities and flare frequencies, the clump mass loss rate and $N_{\mathrm{H}}$ density are highly dependent on the orbital radius considered. If this simple model is correct, the flare frequency\footnote{1 flare every 6 weeks of cumulated \textit{INTEGRAL} observations.} constrains this radius to be of the order of $10^{13}$~cm.
The main difference between SFXTs and persistent super-giant HMXBs would then be their different orbital radius~: a few $R_{*}$ in persistent systems, but tens of $R_{*}$ in SFXTs.

\begin{acknowledgements}
This research has made use of observations from \textit{Swift}, \textit{INTEGRAL}, the ESO/ST-ECF Science Archive, \textit{HST}, 2MASS, of SIMBAD and of IRAF. NM acknowledges the ASI financial support via grant 1/023/05/0. RW thanks Lydia Oskinova for fruitful discussions. JCL acknowledges support through the \textit{XMM-INTEGRAL} PRODEX project and IAP contract P5/36.
\end{acknowledgements}

\bibliographystyle{aa}
\bibliography{Article-Final}

\begin{thebibliography}{26}
\expandafter\ifx\csname natexlab\endcsname\relax\def\natexlab#1{#1}\fi

\bibitem[{{Benensohn} {et~al.}(1997){Benensohn}, {Lamb}, \&
  {Taam}}]{Benensohn+97}
{Benensohn}, J.~S., {Lamb}, D.~Q., \& {Taam}, R.~E. 1997, \apj, 478, 723

\bibitem[{{Brandt} {et~al.}(2006){Brandt}, {Budtz-Joergensen}, {Lund},
  {G\"otz}, {Schanne}, {Rodriguez}, \& {von Kienlin}}]{Brandt+06}
{Brandt}, S., {Budtz-Joergensen}, C., {Lund}, N., {et~al.} 2006, ATel~817

\bibitem[{{Drilling}(1991)}]{Drilling+91}
{Drilling}, J.~S. 1991, \apjs, 76, 1033

\bibitem[{{Feldmeier} {et~al.}(1997){Feldmeier}, {Puls}, \&
  {Pauldrach}}]{Feldmeier+97}
{Feldmeier}, A., {Puls}, J., \& {Pauldrach}, A.~W.~A. 1997, \aap, 322, 878

\bibitem[{{Gehrels} {et~al.}(2004){Gehrels}, {Chincarini}, {Giommi}, {Mason},
  {Nousek}, {Wells}, {White}, {Barthelmy}, {Burrows}, {Cominsky}, {Hurley},
  {Marshall}, {M{\'e}sz{\'a}ros}, {Roming}, {Angelini}, {Barbier}, {Belloni},
  {Campana}, {Caraveo}, {Chester}, {Citterio}, {Cline}, {Cropper}, {Cummings},
  {Dean}, {Feigelson}, {Fenimore}, {Frail}, {Fruchter}, {Garmire}, {Gendreau},
  {Ghisellini}, {Greiner}, {Hill}, {Hunsberger}, {Krimm}, {Kulkarni}, {Kumar},
  {Lebrun}, {Lloyd-Ronning}, {Markwardt}, {Mattson}, {Mushotzky}, {Norris},
  {Osborne}, {Paczynski}, {Palmer}, {Park}, {Parsons}, {Paul}, {Rees},
  {Reynolds}, {Rhoads}, {Sasseen}, {Schaefer}, {Short}, {Smale}, {Smith},
  {Stella}, {Tagliaferri}, {Takahashi}, {Tashiro}, {Townsley}, {Tueller},
  {Turner}, {Vietri}, {Voges}, {Ward}, {Willingale}, {Zerbi}, \&
  {Zhang}}]{Gehrels+04}
{Gehrels}, N., {Chincarini}, G., {Giommi}, P., {et~al.} 2004, \apj, 611, 1005

\bibitem[{{G\"otz} {et~al.}(2006){G\"otz}, {Schanne}, {Rodriguez}, {Leyder},
  {von Kienlin}, {Mowlavi}, \& {Mereghetti}}]{Gotz+06}
{G\"otz}, D., {Schanne}, S., {Rodriguez}, J., {et~al.} 2006, ATel~813

\bibitem[{{Groenewegen} {et~al.}(1989){Groenewegen}, {Lamers}, \&
  {Pauldrach}}]{Groenewegen+89}
{Groenewegen}, M.~A.~T., {Lamers}, H.~J.~G.~L.~M., \& {Pauldrach}, A.~W.~A.
  1989, \aap, 221, 78

\bibitem[{{Horne}(1986)}]{Horne+86}
{Horne}, K. 1986, \pasp, 98, 609

\bibitem[{{Illarionov} \& {Sunyaev}(1975)}]{Illarionov+75}
{Illarionov}, A.~F. \& {Sunyaev}, R.~A. 1975, \aap, 39, 185

\bibitem[{{Kennea} \& {Campana}(2006)}]{Kennea+06}
{Kennea}, J.~A. \& {Campana}, S. 2006, ATel~818

\bibitem[{{Koen} \& {Eyer}(2002)}]{Koen+02}
{Koen}, C. \& {Eyer}, L. 2002, \mnras, 331, 45

\bibitem[{{Krti{\v c}ka} \& {Kub{\'a}t}(2001)}]{Krticka+01}
{Krti{\v c}ka}, J. \& {Kub{\'a}t}, J. 2001, \aap, 377, 175

\bibitem[{{Lamers} {et~al.}(1995){Lamers}, {Snow}, \& {Lindholm}}]{Lamers+95}
{Lamers}, H.~J.~G.~L.~M., {Snow}, T.~P., \& {Lindholm}, D.~M. 1995, \apj, 455,
  269

\bibitem[{{Lang}(1992)}]{Lang+92}
{Lang}, K.~R. 1992, {Astrophysical Data I. Planets and Stars.} (937 pp.~33
  figs..~ Springer-Verlag Berlin Heidelberg New York)

\bibitem[{{Leitch} \& {Vasisht}(1998)}]{Leitch+98}
{Leitch}, E.~M. \& {Vasisht}, G. 1998, New Astronomy, 3, 51

\bibitem[{{Masetti} {et~al.}(2006){Masetti}, {Bassani}, {Bazzano}, {Dean},
  {Stephen}, \& {Walter}}]{Masetti+06ATel}
{Masetti}, N., {Bassani}, L., {Bazzano}, A., {et~al.} 2006, ATel~815

\bibitem[{{Mereghetti} {et~al.}(2006){Mereghetti}, {Sidoli}, {Paizis}, \&
  {G\"otz}}]{Mereghetti+06}
{Mereghetti}, S., {Sidoli}, L., {Paizis}, A., \& {G\"otz}, D. 2006, ATel~814

\bibitem[{{Perryman} {et~al.}(1997){Perryman}, {Lindegren}, {Kovalevsky},
  {Hoeg}, {Bastian}, {Bernacca}, {Cr{\'e}z{\'e}}, {Donati}, {Grenon}, {van
  Leeuwen}, {van der Marel}, {Mignard}, {Murray}, {Le Poole}, {Schrijver},
  {Turon}, {Arenou}, {Froeschl{\'e}}, \& {Petersen}}]{Perryman+97}
{Perryman}, M.~A.~C., {Lindegren}, L., {Kovalevsky}, J., {et~al.} 1997, \aap,
  323, L49

\bibitem[{{Remillard} {et~al.}(2006){Remillard}, {Lin}, {Cooper}, \&
  {Narayan}}]{Remillard+06}
{Remillard}, R.~A., {Lin}, D., {Cooper}, R.~L., \& {Narayan}, R. 2006, \apj,
  646, 407

\bibitem[{{Schlegel} {et~al.}(1998){Schlegel}, {Finkbeiner}, \&
  {Davis}}]{Schlegel+98}
{Schlegel}, D.~J., {Finkbeiner}, D.~P., \& {Davis}, M. 1998, \apj, 500, 525

\bibitem[{{Sguera} {et~al.}(2006){Sguera}, {Bazzano}, {Bird}, {Dean},
  {Ubertini}, {Barlow}, {Bassani}, {Clark}, {Hill}, {Malizia}, {Molina}, \&
  {Stephen}}]{Sguera+06}
{Sguera}, V., {Bazzano}, A., {Bird}, A.~J., {et~al.} 2006, \apj, 646, 452

\bibitem[{{Skinner} {et~al.}(1982){Skinner}, {Bedford}, {Elsner}, {Leahy},
  {Weisskopf}, \& {Grindlay}}]{Skinner+82}
{Skinner}, G.~K., {Bedford}, D.~K., {Elsner}, R.~F., {et~al.} 1982, \nat, 297,
  568

\bibitem[{{Snow} {et~al.}(1994){Snow}, {Lamers}, {Lindholm}, \&
  {Odell}}]{Snow+94}
{Snow}, T.~P., {Lamers}, H.~J.~G.~L.~M., {Lindholm}, D.~M., \& {Odell}, A.~P.
  1994, \apjs, 95, 163

\bibitem[{{van Genderen} {et~al.}(1989){van Genderen}, {Bovenschen},
  {Engelsman}, {Goudfrooy}, {van Haarlem}, {Hartmann}, {Latour}, {Ng}, {Prein},
  {van Roermund}, {Roogering}, {Steeman}, \& {Tijdhof}}]{vanGenderen+89}
{van Genderen}, A.~M., {Bovenschen}, H., {Engelsman}, E.~C., {et~al.} 1989,
  \aaps, 79, 263

\bibitem[{{Walborn}(1973)}]{Walborn+73}
{Walborn}, N.~R. 1973, \aj, 78, 1067

\bibitem[{{Winkler} {et~al.}(2003){Winkler}, {Courvoisier}, {Di Cocco},
  {Gehrels}, {Gim{\' e}nez}, {Grebenev}, {Hermsen}, {Mas-Hesse}, {Lebrun},
  {Lund}, {Palumbo}, {Paul}, {Roques}, {Schnopper}, {Sch{\" o}nfelder},
  {Sunyaev}, {Teegarden}, {Ubertini}, {Vedrenne}, \& {Dean}}]{WInkler+03}
{Winkler}, C., {Courvoisier}, T.~J.-L., {Di Cocco}, G., {et~al.} 2003, \aap,
  411, L1

\end{thebibliography}

\newpage

\end{document}